\documentclass[final,3p,twocolumn,authoryear,hidelinks]{elsarticle}
\usepackage{graphicx}
\usepackage{amssymb}
\usepackage{amsmath}
\usepackage{tabu}
\usepackage{longtable}
\usepackage[table]{xcolor}
\usepackage{afterpage}
\usepackage[hyphens]{url}
\usepackage[unicode,pdfhighlight=/P,colorlinks=true]{hyperref}

\newcommand{\Geo}[2]{\ensuremath{#1^\circ\,\text{#2}}}
\newcommand{\E}[1]{\Geo{#1}{E}}

\newcommand{\N}[1]{\Geo{#1}{N}}
\newcommand{\isotope}[2]{\ensuremath{{}^{\text{#1}}\text{#2}}}
\newcommand{\Cs}[1]{\isotope{#1}{Cs}}

\begin{document}
\hypersetup{colorlinks=true,linkcolor=red,citecolor=brown,urlcolor=blue,unicode,pdfhighlight=/P}
\begin{frontmatter}

%% Title, authors and addresses
%% use the tnoteref command within \title for footnotes;
%% use the tnotetext command for the associated footnote;
%% use the fnref command within \author or \address for footnotes;
%% use the fntext command for the associated footnote;
%% use the corref command within \author for corresponding author footnotes;
%% use the cortext command for the associated footnote;
%% use the ead command for the email address,
%% and the form \ead[url] for the home page:
%%
%% \title{Title\tnoteref{label1}}
%% \tnotetext[label1]{}
%% \author{Name\corref{cor1}\fnref{label2}}
%% \ead{email address}
%% \ead[url]{home page}
%% \fntext[label2]{}
%% \cortext[cor1]{}
%% \address{Address\fnref{label3}}
%% \fntext[label3]{}

\title{Role of mesoscale eddies in transport of Fukushima-derived cesium
isotopes in the ocean}

%% use optional labels to link authors explicitly to addresses:
%% \author[label1,label2]{<author name>}
%% \address[label1]{<address>}
%% \address[label2]{<address>}
\author{M.V.~Budyansky}
\author{V.A.~Goryachev}
\author{D.D.~Kaplunenko}
\author{V.B.~Lobanov}
\author{S.V.~Prants}
\ead{prants@poi.dvo.ru}
\author{A.F.~Sergeev}
\author{N.V.~Shlyk}
\author{M.Yu.~Uleysky}
%\ead{prants@poi.dvo.ru}
%\ead[url]{www.dynalab.poi.dvo.ru}

\address{V.I.Il'ichev Pacific Oceanological Institute of the 
Far-Eastern Branch of the Russian Academy of Sciences,\\
43 Baltiyskaya st., 690041 Vladivostok, Russia\\
URL: \url{http://dynalab.poi.dvo.ru}}

\begin{abstract}
%We present the results of measurement of \Cs{134} and \Cs{137} released
%from the Fukushima Nuclear Power Plant (FNPP) 
%in seawater samples collected in the cruise 
%%of R/V ``Professor Gagarinskiy'' 
%at surface and different depths in the western North Pacific in June and July 2012.

We present the results of {\it in-situ} measurements of \Cs{134} and \Cs{137} 
released from the Fukushima Nuclear Power Plant (FNPP) collected at surface 
and different depths in the western North Pacific in June and July 2012. 
It was found that 15 month after the incident concentrations of radiocesium in 
the Japan and Okhotsk seas were at background or slightly increased level, 
while they  
had increased values in the subarctic front area east of Japan. 
The highest concentrations of \Cs{134} and \Cs{137} up to 13.5 ${\pm}$ 0.9 
and 22.7 ${\pm}$ 1.5 Bq~m$^{-3}$ 
have been found to exceed ten times the background levels before the accident. 
Maximal content of radiocesium was observed within subsurface and intermediate 
water layers inside the cores of anticyclonic eddies (100 -- 500~m). Even slightly increased content of radiocesium 
was found at some eddies at depth of 1000~m. It is expected that convergence and subduction 
of surface water inside eddies are main mechanisms of downward transport of radionuclides. 
{\it In situ} observations are 
compared with the results of simulated advection of these radioisotopes by the AVISO altimetric 
velocity field. Different Lagrangian diagnostics are used to reconstruct the history and 
origin of synthetic tracers imitating measured seawater samples collected in each of those eddies. 
The results of observations are consistent with the simulated results. 
It is shown that the tracers, simulating water samples with increased radioactivity 
to be measured in the cruise, really visited the areas with presumably high level of 
contamination. Fast water advection between anticyclonic eddies and convergence 
of surface water inside eddies make them responsible for spreading, accumulation 
and downward transport of cesium rich water to the intermediate depth in the frontal zone. 
\end{abstract}
\begin{keyword}
%% keywords here, in the form: keyword \sep keyword
Fukushima accident \sep radiocesium observation \sep Lagrangian modelling
\sep mesoscale eddies
%\sep  Kuroshio rings
%% MSC codes here, in the form: \MSC code \sep code
%% or \MSC[2008] code \sep code (2000 is the default)
\end{keyword}
\end{frontmatter}

\section{Introduction}

The great Tohoku earthquake of magnitude 9.0 on 11 March 2011
followed by the tsunami
inflicted heavy damage on the Fukushima Nuclear Power Plant (FNPP) due to
overheating the reactors and hydrogen explosions.
Large amount of radioactive water leaked directly into the ocean
\citep{Tsumune2012,Kanda2013}. Moreover,
the radioactive pollution of the sea was also caused by atmospheric deposition on
the ocean surface \citep{Takemura2011,Miyazawa2012}.
Radioactive cesium with 30.17~yr half-life for
\Cs{137} and 2.06~yr half-life for \Cs{134} has been detected
over a broad area in the North Pacific in 2011 and 2012
\citep{Honda12,Buesseler12,Inoue2012a,Inoue2012,Kaeriyama13,Oikawa2013,Aoyama2013,Kamenik2013,
Kumamoto2014,Kaeriyama2014}.
\Cs{137} is a passive tracer in seawater which can be used
to study long-term circulation and ventilation of water masses in the global ocean. In particular,
distribution of Fukushima-derived \Cs{137} in the ocean would help to validate
numerical circulation models and their parameters.

The area east of Japan is known as Kuroshio\,--\,Oyashio confluence zone \citep{Kawai1972}
or subarctic front area. Existence of a large number of mesoscale eddies in this area
should influence the transport of water contaminated with Fukushima-derived radionuclides.
Among these eddies, the Kuroshio warm core rings are most energetic and long-lived ones
\citep[see, e.g.,][]{Kitano1974,Itoh10}.
Strong and persistent anticyclonic eddies are also observed along the Kuril Islands
\citep{Bulatov1992,Yasuda2000}. 

Both cyclonic and anticyclonic eddies would provide fast transport of surface water 
by streamers. However, because of a divergence in the cyclonic eddies one may expect 
upwelling of deep water and thus lower concentration of radionuclides in the surface 
layer. In opposite, we may expect accumulation of contaminated
water in anticyclonic eddies because of convergence in their surface layer and following
subduction to the deeper ocean. In addition, winter convection should increase penetration of
contaminated water to deeper layers inside anticyclonic eddies in comparison with surrounding
waters. 

It is known that Kuroshio warm-core rings may move north-eastward toward the Kuril Islands
\citep{Lobanov1991,Bulatov1992,Yasuda92,Itoh10} and thus transport trapped
water with a higher cesium content to the north of the subarctic front. To prove this hypothesis we
have performed a numerical modelling of tracer transport in the area east of Japan and
implemented a cruise to cross major streams and eddies in the area using R/V
``Professor Gagarinskiy'' of the Far Eastern Branch of Russian Academy of Sciences.
The cruise has been conducted from 12~June to 10~July 2012, 15 months after the accident.

We focus here on comparing the experimental results, obtained with water
samples at stations in the centers of some selected anticyclonic eddies in the region,
with the results of numerical simulation of spatial distribution of Fukushima-derived
radionuclides. That simulation helps to explain why we have detected comparatively high cesium
concentrations in the cores of some eddies and lower ones in the other eddies.

The paper is organized as follows. In section~2 we briefly present for comparison the results of
previous direct observations of \Cs{134} and \Cs{137} in the area of the
western North Pacific \citep{Honda12,Buesseler12,Kaeriyama13} where some of our sampling stations
were located.
Section~3 describes the cruise track, sample collection, the methods of
radioactive analysis and computation. The numerical Lagrangian
methods we used are based on solving advection equations
for synthetic particles in an altimetric velocity field provided by AVISO.
Section~4 contains the main results and a discussion. We present a table summarizing the results of
our measurements of the concentration of \Cs{134} and \Cs{137} in seawater samples
collected at different depth horizons in the broad area in the Sea of Japan, the Okhotsk Sea and
the western North Pacific. Simulation results are plotted as tracking maps
revealing origin and history of water masses collected at representative sampling stations where
different values of concentration of \Cs{134} and \Cs{137} have been observed.
They are compared with the corresponding measurements. In this section we analyze as well
vertical cross-section of potential density anomaly, potential vorticity, distribution of cesium
isotopes in surface water along the cruise track and vertical distribution of \Cs{134}
and \Cs{137} at some selected stations. It is also discussed why measured activities
of cesium differs strongly at different stations. Results are summarized in Section 5.

\section{Previous observations of Fukushima-derived cesium}

Before March 2011, \Cs{137} concentration levels off Japan were 
1--2~Bq~m$^{-3}$ ${\simeq}$
0.001--0.002~Bq~kg$^{-1}$, while \Cs{134} was not detectable.
Because of a comparatively short half-life time, any measured concentrations of
\Cs{134} could only be Fukushima derived.
Concentrations at the FNPP discharge channels in early April 2011 was more than 50 million
times greater than the preexisting ocean level of \Cs{137} \citep{Buesseler12}.

One month after the accident, sea-water, suspended solids and
zooplankton samples were collected from the surface mixed layer and subsurface
layers at a number of stations, 200--2000~km offshore from the FNPP \citep{Honda12}.
In surface water, \Cs{137} concentrations were ranged from several times to two orders of
magnitude higher than before the accident. \Cs{134} isotope was also detected
with the ratio $\Cs{134}/\Cs{137}$ to be about~1. The highest
concentrations, from ${\approx} 150$~Bq~m$^{-3}$  to ${\approx} 350$~Bq~m$^{-3}$,
have been found off the FNPP (${\approx} 200$~km from the nuclear power plant) and
Miyagi (the earthquake source). \Cs{137} concentrations to the east of this
region in the area (\E{146\text{--}147}; \N{37\text{--}38}) were also high
(${\approx}50$--$60$~Bq~m$^{-3}$). The \Cs{137}
concentrations in the Kuroshio Extension, ${<}10$~Bq~m$^{-3}$, were unexpectantly low,
because it was considered to be the main potential pathway for contaminated water
to the open ocean.

The expedition of the Russian Hydrometeorological Service on R/V ``Pavel Gordienko''
in 24 April -- 6 May 2011 \citep{Karasev2012} proved increased concentration 
of both \Cs{137} and \Cs{134} in surface water along the whole Kuril Island chain
(2.2--3.6 Bq~m$^{-3}$ and 1.2--2.9 Bq~m$^{-3}$, correspondingly) 
but not at the southern part of the Kamchatka Peninsula, where those 
concentrations were 1.4 and 0.4 Bq~m$^{-3}$, correspondingly. Such distribution of radionuclides could be explained
by atmospheric transport. They also found high concentration of radionuclides in the area
about 350~km east of Tohoku, where \Cs{137}  and \Cs{134}  contents were
found to be up to 24 and 21 Bq~m$^{-3}$, correspondingly.

The R/V ``Ka'imikai-o-Kanaloa'' cruise has been conducted in 4--18 June 2011
by \citep{Buesseler12} to investigate the distribution of Fukushima-derived radionuclides in
seawater, zooplankton and micronectonic fishes
30--600~km offshore from the FNPP. Activities up to $325$~Bq~m$^{-3}$ 
were found more than 600~km offshore.
As to \Cs{137}, the highest level (except for the
discharge channels), 600 -- 800~Bq~m$^{-3}$, has been detected 30~km offshore.
In June, Fukushima-derived \Cs{} did not generally penetrate below 100--200~m.
Over time, it is expected to find deeper penetration proving a means to
study the rates of vertical mixing processes in the Pacific.  Fukushima-derived
isotopes  have been also detected in zooplankton (with the maximal level
about $5\cdot 10^4$~Bq~m$^{-3}$  dry weight comparable with
the recommended value of~$4 \cdot 10^4$) and jellyfish but not in micronectonic fishes.
In June 2011, the highest surface-water
concentrations for both the isotopes, $3.9 \cdot 10^3$~Bq~m$^{-3}$, have been detected
in a semipermanent mesoscale eddy centered at (\E{142.5}; \N{37}),
not the nearest location to the nuclear power plant. 

Results of direct observation of \Cs{134} and \Cs{137} in surface seawater collected
from R/V ``Kaiun maru'' in a broad area in
the western and central North Pacific in July, October 2011 and July 2012 have been reported
by~\citet{Kaeriyama13}.
In particular, seawater samples were collected at their stations C43--C55
(26--29 July 2011) located from  \N{35} to \N{41} along the \E{144} transect
with its southern edge crossing a crest of the Kuroshio Extension meander
and the northern edge crossing partly the Tohoku mesoscale eddy centered at that time at
(${\approx}\E{144}$; \N{38}). That eddy is
a warm-core Kuroshio ring permanently present in the region till the end of our cruise and later.
It is clearly seen in an earlier simulation of Fukushima-derived radionuclides propagation
\citep[see Fig.~3b by][]{DAN11} and in the present one marked by letter 'T' 
%as a patch with increased density of tracers 
on the Lagrangian maps in Fig.~\ref{fig3}. During 15 months after the accident, that eddy
has interacted with a number of adjacent eddies and streamers promoting transport of contaminant
water to the north, south and east. The measured \Cs{137}
concentrations at stations C43--C55 have been varied from the background level of
$1.9\pm 0.4$~Bq~m$^{-3}$  (station C52) to $153\pm 6.8$~Bq~m$^{-3}$ (station C47).
The ratio $\Cs{134}/\Cs{137}$ was close to~1.

\section{Materials and methods}
\begin{figure}[!htb]
\centerline{\includegraphics[width=0.45\textwidth,clip]{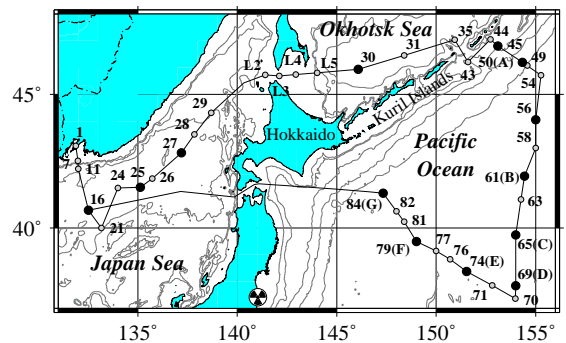}}
\caption{Locations and numbers of stations where surface (open circles) and deep seawater 
samples (full circles) 
were collected during the cruise (12~June -- 10~July 2012). Letters A, B, C, D, E, F and G 
mark elliptic points in the centers of the corresponding mesoscale eddies to be studied. 
Radioactivity sign is location of the FNPP.}
\label{fig1}
\end{figure}
%

%\afterpage{\include{tab}}

The cruise on the board of  R/V ``Professor Gagarinskiy'' was conducted with the aim to
collect data on the  distribution of artificial radionuclides after the accident
at the FNPP in the area of the Japan Sea, Sea of Okhotsk and the adjacent area of the Northwest 
Pacific (Fig.~\ref{fig1}).
We used standard methods to collect water and biota
followed by laboratory processing and detection of \Cs{134} and \Cs{137}
with a high-purity germanium spectrometer.

\subsection{Sample collection}

During the cruise, conducted from 12~June to 10~July 2012, surface water samples were collected
along the cruise track by a submerged pump at 54 stations. Water samples from subsurface and
deep layers (100--3621~m) were taken using a CTD/Rosette sampling system at 15 stations.
A volume of each water sample was 95--120~liters. Locations of stations are shown in Fig.~\ref{fig1}.
The water samples pre-treatment for measurement of \Cs{134} and \Cs{137} isotopes
was performed on board by concentrating the cesium isotopes on the selective sorbent ANFEZH
\citep{Remez,Remez1996,Bandong2001} with a preliminary separation
of suspended matter. Water was pumped through a filter and then through the sorbent with
a flow rate around 2~liters per minute.
\begin{figure}[!htb]
\centerline{\includegraphics[width=0.45\textwidth,clip]{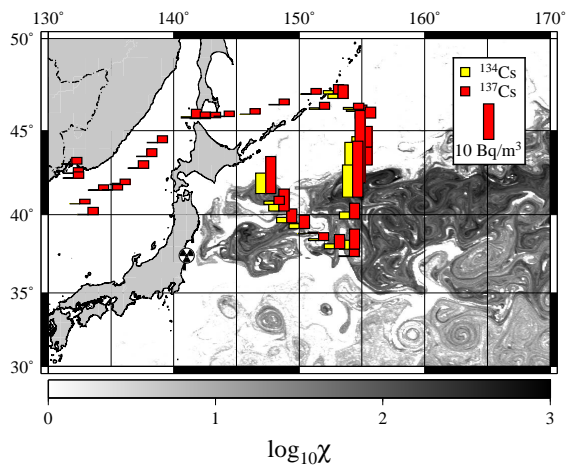}}
\caption{Simulated distribution of radionuclides concentration in the Northwest
Pacific to the end of June 2012 with measured concentrations
of \Cs{134} and \Cs{137} in Bq~m$^{-3}$ imposed. The relative simulated concentration,
$\chi$, is in a logarithmic scale.}
\label{fig2}
\end{figure}

\subsection{Measurements of \Cs{134} and \Cs{137} in seawater}
\begin{figure*}[!htb]
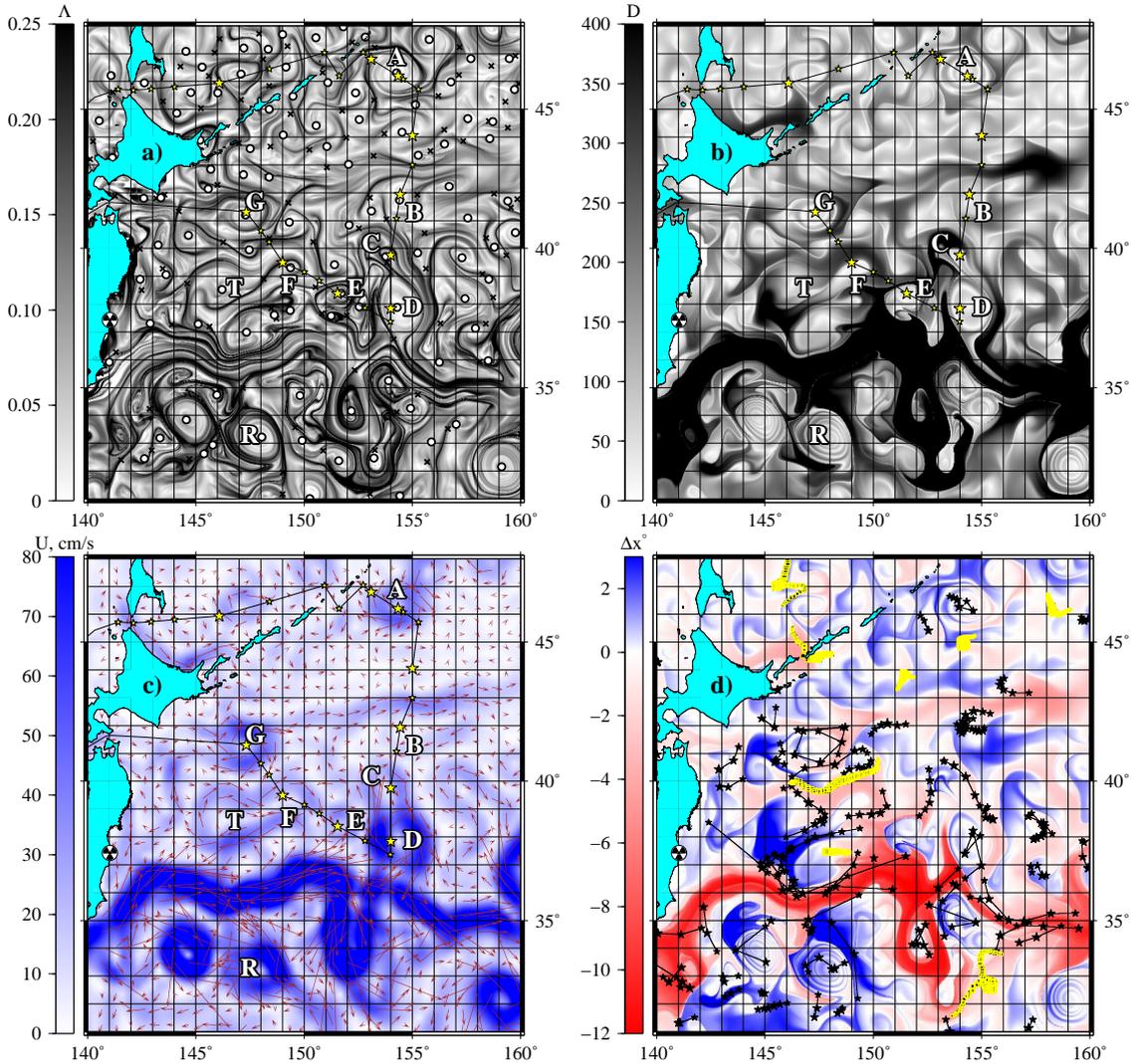

\centerline{\includegraphics[width=0.45\textwidth,clip]{fig3a.eps}
\includegraphics[width=0.45\textwidth,clip]{fig3b.eps}}
\centerline{\includegraphics[width=0.45\textwidth,clip]{fig3c.eps}
\includegraphics[width=0.45\textwidth,clip]{fig3d.eps}}
\caption{Anticyclonic eddies of the subarctic front
A, B, C, D, E, F and G (a) on the Lyapunov map ($\Lambda$ is in days$^{-1}$), 
(b) on the drift map (the absolute displacement of tracers $D$ is in km) and 
(c) in the altimetric velocity field (the speed $U$ is in cm/sec) with elliptic (circles) and hyperbolic
(crosses) ``instantaneous'' stagnation points imposed on 28 June 2012. 
The ship's track and some sampling stations are shown. 
(d) Backward-in-time zonal drift map on 28 June 2012 with ``light'' (``red'' online) 
and ``dark'' (``blue'' online)
waters passing for 15 days large distances in the west\,--\,east and the east\,--\,west
directions, respectively (for interpretation of the colors in this figure, the reader is referred to
the web version of this article). The values of zonal displacements of
particles, $\Delta x \equiv x_f - x_0$, are given in geographic degrees. Tracks of buoys
ARGOS (stars) and drifters (squares) are shown in the area for the same period of time.}
\label{fig3}
\end{figure*}

Further processing of the samples and measurements of gamma activity have been continued at
the land-based laboratory at the V.I.~Il`ichev Pacific Oceanological Institute in Vladivostok.
Each sorbent has been dried in the oven for 3--5~hours at temperatures of 70--80~$^{\circ}$C
and then burned in a muffle furnace at temperatures of 430--450~$^{\circ}$C during 10--15~hours.
Initial sorbent of each water sample had a volume of 320~cm$^3$ and a weight of 80~g, while
the ash in the measuring vessel for gamma spectrometry, remained after combustion of the sorbent,
had a volume of 4.5--4.7~cm$^3$. Minimizing the sample volume is essential for gamma-spectrometric
analysis as it allows decreasing a minimum detective activity.
Recovery efficiency of cesium and \isotope{60}{Co} was $0.98 \pm 0.02$ and $0.67 \pm 0.1$, respectively,
where errors $\pm 0.02$ and $\pm 0.1$ are equal to $2\sigma$ of ten \Cs{137} and \isotope{60}{Co}
labeled sea water samples processing.

Gamma activity and radioisotope composition have been determined with a gamma spectrometer with a high
purity germanium detector GEM150 and a digital multi-channel analyzer DSPEC jr 2.0 (ORTEC).
To reduce the background activity, the detector was placed in a lead shield with 10~cm thickness
of the wall and cover and inner walls covered with a copper layer of 1 mm thickness.
Integral background count rate of the detector in the shield within the energy range
of 50--2990~KeV is 6.6 counts per second. We have obtained the following minimum detective activities
for the measurement of 80~g ANFEZH blank sample during 160000~s:
\Cs{134}~--- 0.004~Bq,
\Cs{137}~--- 0.005~Bq and \isotope{60}{Co}~--- 0.014~Bq. For the spectrometer calibration we used an 80~g
sorbent soaked with solution of known concentration of \Cs{134}, \Cs{137} and \isotope{60}{Co} after
its ashing. The gamma-ray spectrum was analyzed using the software package Gamma Vision-32,
version 6.09, ORTEC. True-coincidence correction factor for \Cs{134} in our case was 
determined as 1.64. The \Cs{134} to \Cs{137} ratio was found to be 0.62 
(Fig.~1s in supplementary material), 
whereas at the time of accident this value was equal 1. 
This decrease corresponds to a partial disintegration of \Cs{134} for 15 months after the accident. 
All results were corrected to the sampling date.

\subsection{Lagrangian simulation}

To simulate propagation of Fukushima-derived radionuclides we apply
in this paper the Lagrangian approach that is based on computing trajectories of tracers
advected by an AVISO velocity field in accordance with the following equations:
\begin{equation}
\frac{d x}{d t}= u(x,y,t),\quad \frac{d y}{d t}= v(x,y,t),
\label{adveq}
\end{equation}
where $x$ and $y$ are the longitude and latitude of a tracer in geographical
minutes, $u$ and $v$ are angular zonal and meridional components
of the surface velocity expressed in minutes per day.
Geostrophic velocity field, from the day of accident 11 March 2011  
to the end of the cruise in the Pacific ocean area on 5 July 2012 (station 84),  
were obtained from the AVISO database (\url{aviso.oceanobs.com}).
The data is gridded on a $1/3^\circ\times1/3^\circ$ Mercator grid.
Bicubical spatial interpolation and third order Lagrangian polynomials in time have been used
to provide accurate numerical results.

\begin{figure}[!htb]
\centerline{\includegraphics[width=0.45\textwidth,clip]{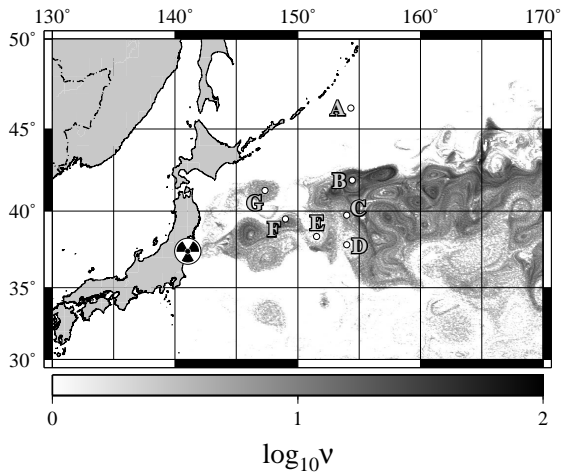}}
\caption{Simulated forward-in-time tracking map for the ``radioactive'' tracers distributed
initially inside the box around the FNPP with the coordinates
(\E{140\text{--}144}; \N{36.5\text{--}38.5}). The map shows where those tracers have been 
from 24 June to 6 July 2012. Letters A, B, C, D, E, F and G 
mark elliptic points (circles) in the centers of the corresponding mesoscale eddies to be studied. 
The density of tracers, $\nu$, is in a logarithmic scale.}
\label{fig4}
\end{figure}

When integrating Eqs.~(\ref{adveq}) forward
in time, one gets an information about the fate of tracers. That is a common way
to simulate the horizontal distribution pattern of radionuclides released from the FNPP
\citep{KAWAMURA2011,Dianskii2012,Nakano2012,Dietze2012,Miyazawa2012,Behrens2012,Rypina2013,
Tsumune2013,Choi2013,Rossi2013,Maderich2014}. We present forward-in-time simulation results in  
Figs.~\ref{fig2} and \ref{fig4}. In order to compute Lyapunov and other Lagrangian maps  
with eddies and streamers to be present in the area studied, we perform backward-in-time 
integration for 15 days. The corresponding results are shown in Fig.~\ref{fig3}. 
The Lagrangian maps have been shown to be useful in studying transport and mixing
in different basins, from marine bays \citep{FAO13} and seas \citep{OM11}
to the ocean scale \citep{DAN11,P13}. 

Our main goal in simulation is to trace out the origin and history of water samples
that have been collected 15 months and later after the accident to measure the
cesium concentration. To do that we seed a square $2 \times 2$~km around the location 
of a given sampling station with tracers and advect them backward in time in 
the altimetric velocity field, beginning from the date of sampling to the day of the accident. 
Fixing the places on a regional map,
where the corresponding artificial tracers have been found for one month after the accident,
we can estimate by the tracer density the probability to detect higher
concentrations of Fukushima-derived radionuclides in surface seawater samples at sampling  
stations. The corresponding backward-in-time tracking maps aree shown in Figs.~\ref{fig8}, \ref{fig9} 
and \ref{fig10}. The simulated results are compared with 
the results of measurements of \Cs{134} and \Cs{137}.
%at the same stations.
%

The specific Lagrangian maps are used to identify eddies and transport pathways in
the region. The finite-time Lyapunov exponent (FTLE) field is known to
characterize quantitatively mixing of water \citep{Pierrehumbert}
and can be used, in particular, to identify eddies, streamers and fronts in irregular velocity fields
in the area \citep{DAN12,P13,Prants2014,Prants2014c}.
The FTLE is computed here by the method
of the singular-value decomposition of an evolution matrix for the linearized
advection equations \citep{OM11} with the help of the formulae   
\begin{equation}
\Lambda(t,t_0)=\frac{\ln\sigma(t,t_0)}{t-t_0},
%\label{lyap_ftle}
\end{equation}
which is the ratio of the logarithm of the maximal possible
stretching in a given direction to the integration time interval $t-t_0$.
Here $\sigma(t,t_0)$ is the maximal singular value of the evolution matrix.
This method enables to identify accurately mesoscale eddies in altimetric velocity fields.
Another means to identify eddies and transport pathways is provided by the
absolute, zonal and meridional drift maps \citep{DAN11,DAN12,FAO13,Prants2014,Prants2014c}.
The finite-time absolute displacement is a distance between
final and initial positions of advected particles on the sea surface.
We compute the absolute displacement and its zonal
and meridional components backward in time for a large number of particles in order to
identify mesoscale eddies and zonal and meridional jets to be present in the region for
a given period of time. The absolute displacement, $D$, is simply a distance between 
final ($x_f, y_f$) and initial ($x_0, y_0$) positions of advected particles on the Earth sphere 
with the radius $R$
\begin{equation}
D\equiv R\arccos[\sin y_0 \sin y_f +\cos y_0 \cos y_f \cos (x_f - x_0)].
\label{drift}
\end{equation}

Stagnation points, where the altimetric velocity is found to be zero, 
contain important information about the regional flow. We compute them daily, check 
their stability type by standard stability tests and impose them on FTLE and drift 
maps as circles (elliptic points) and crosses (hyperbolic points).
The elliptic points, situated mainly in the centers of eddies, are those ``instantaneous'' 
stagnation points around which the motion is stable and circular.
The hyperbolic points, situated mainly between and around of eddies, are unstable ones 
with two directions along which waters converge to such a point and another two directions
along which they diverge.  

\section{Results and discussion}

In this section we present the results of direct observations of
\Cs{134} and \Cs{137} in surface seawater and at different depths
in a broad area in the Sea of Japan, the Okhotsk Sea and
the western North Pacific and discuss measured levels of
the isotope concentrations. Those observations are compared with the
results of numerical simulation
of spatial distribution of Fukushima-derived radionuclides based on
the altimetric velocity field.

%\subsection{Simulation of lateral distribution patterns of Fukushima-derived cesium}
\subsection{Observed and simulated horizontal distribution patters of radiocesium}

Concentrations of \Cs{137} in the Japan and Okhotsk Seas have been found to be  
1.4--2.3 and 1.5--1.9~Bq~m$^{-3}$, 
accordingly (Table~1 in Appendix and Fig.~\ref{fig2}), and did not exceed much pre-accident level. A slightly increased concentration
of \Cs{134}  (2.4~Bq~m$^{-3}$) was found only at one station L2 located in the northeastern
Japan Sea
off northern tip of the Hokkaido Island. This station was sampled in the area of the Tsushima Current
which transports water contaminated by river runoff from the Honshu Island that could explain
a higher concentration of \Cs{137}. This is in accordance with observations by
\citep{Inoue2012} which showed an increase of \Cs{134} and \Cs{137} concentrations in surface
water transported by the Tsushima Current along the west coast of Japan to Hokkaido.

All surface water samples, collected in the Pacific Ocean, contain an increased concentration of
\Cs{134}, 0.2--11.9~Bq~m$^{-3}$, with except of Station 76 (Table~1 in Appendix and Fig.~\ref{fig2}). 
This station was taken in the warm streamer extended
northward from the Kuroshio Extension jet and thus transported relatively clean water. \Cs{137}
concentrations in all samples in the western subarctic Pacific and Kuroshio\,--\,Oyashio frontal zone
were in the range of 1.8--21~Bq~m$^{-3}$. The surface water maximal \Cs{134}  and \Cs{137}
concentrations were registered in the frontal zone between stations 56--65 and stations 81--84 
(Fig.~\ref{fig2}).
This confirms accumulation of surface water in the frontal zone with particular increased content
in anticyclonic mesoscale eddies. Below we will discuss a role of those eddies in more details.

Solving the advection equations (\ref{adveq})
backward in time, we have computed FTLE and drift Lagrangian maps on
each cruise day. Those maps have been sent electronically to the board and
used to correct the cruise track in order to cross prominent eddies
in the area. Below we show simulated lateral distribution of radionuclides and compare 
that with cruise observations.

The initial distribution of radionuclides is supposed to be a patch with
tracer concentration decreasing logarithmically with distance from the FNPP location.
We advect tracers
by the corresponding altimetric velocity field, starting not from the date of tsunami but 
from 25 March 2011, 
in order to take into account not only a direct release of radioactive material 
from the FNPP but, as well, a subsequent atmospheric deposition on the ocean surface 
just after the tsunami on 11 March. 
Variations in the initial date do not change significantly the simulation results.
The concentration distribution, $\chi$, in the end of June 2012 is shown
in Fig.~\ref{fig2} in a logarithmic scale with measured concentrations of \Cs{134} and \Cs{137}
imposed. 

As expected, tracers in the surface layer were transported mainly along
the Kuroshio Extension to the east. The concentration is larger on the north
flank of the parent jet because the larger part of the initial radioactive patch
was situated to the north from the eastward jet. Transport of radionuclides
to the southern flank of the Kuroshio Extension may be explained partly by tracer
advection from the southern part of the initial patch. Moreover, there exist
another transport pathways for Fukushima-derived radionuclides. One of them, a
cross-jet transport, have been studied numerically in the AVISO field
by \citep{Prants2014} and confirmed by tracks of surface drifters released during the
``Ka'imikai-o-Kanaloa'' cruise \citep{Buesseler12} and before.
The mechanism of meridional cross-jet transport, documented by \citep{Prants2014},
is pinching off rings with contaminated water from the southern flank of the Kuroshio Extension jet.
We have not found the significant
impact of the initial patch's size on the concentration distribution.
{\it In situ} observations are consistent qualitatively with the simulation:
the maximal measured concentrations at the stations selected have been
detected in those areas in Fig.~\ref{fig2} where the density of artificial tracers
is really comparatively high.

\subsection{Distribution of anticyclonic mesoscale eddies}
\begin{figure}[!htb]
\centerline{\includegraphics[width=0.45\textwidth,clip]{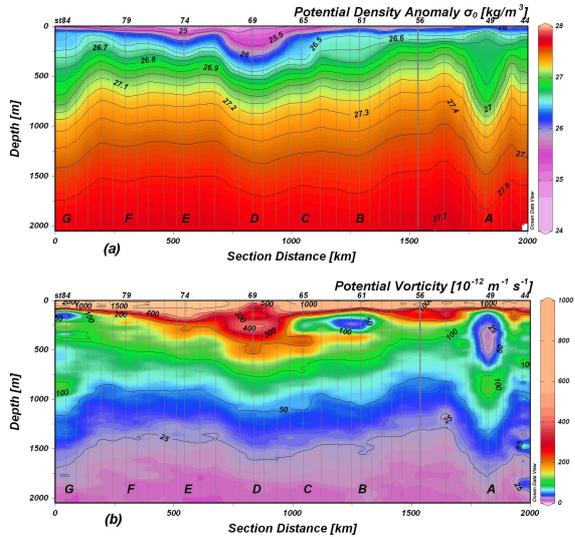}}
\caption{Vertical cross-section of (a) potential density anomaly and (b) potential vorticity
along the cruise track from the central Kuril Islands (right) to Hokkaido (left).
Station numbers are indicated along the top axis, locations of anticyclonic 
eddy centers are marked as A\,--\,G.}
\label{fig5}
\end{figure}
\begin{figure}[!htb]
\centerline{\includegraphics[width=0.45\textwidth,clip]{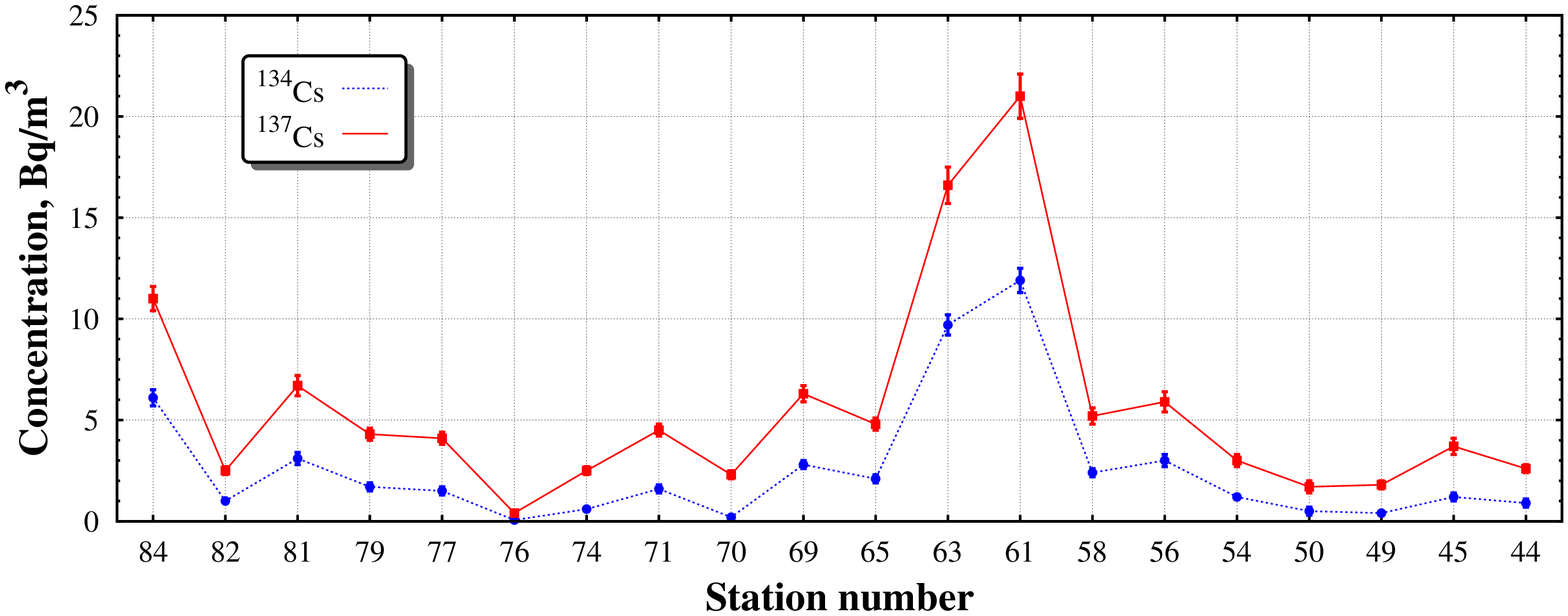}}
\caption{Distribution of cesium isotopes \Cs{134} (dotted line) and \Cs{137}
in surface water along the cruise track in the Pacific.
Locations of stations, centers of the mesoscale eddies to be studied 
and cruise track are shown in Fig.~\ref{fig1}.}
\label{fig6}
\end{figure}
\begin{figure}[!htb]
\centerline{\includegraphics[width=0.45\textwidth,clip]{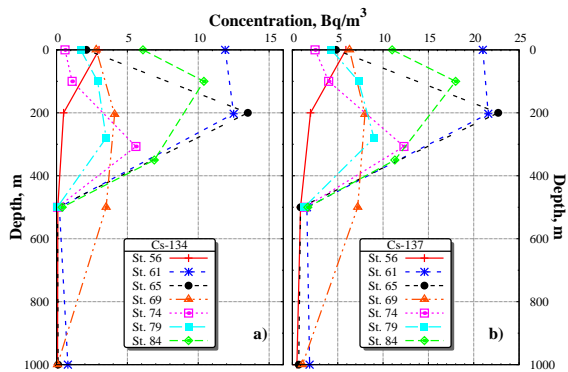}}
\caption{Vertical distribution of (a) \Cs{134} and (b) \Cs{137}
in Bq m$^{-3}$ for some selected stations.}
\label{fig7}
\end{figure}

The cruise track, shown on the FTLE map in Fig.~\ref{fig3}a, b and c was chosen to cross the eddies A, B, C, D, E, F and G
marked on that figure. All they are anticyclonic mesoscale eddies of the subarctic front
but of different origin and history. The eddy A is a long-lived quasi-stationary Kuril anticyclone.
The eddy B appeared in the northern subarctic front area on the southern
flank of a zonal eastward jet transporting waters along \N{42\text{--}43} from the eastern coast of
Japan (see Fig.~\ref{fig3}d). The eddy C forms a pair along with the warm-core Kuroshio
ring D that was pinched off from a Kuroshio Extension meander in the end of May 2012.
Both are clearly seen on the velocity map in Fig.~\ref{fig3}b
as a vortex pair. The warm-core Kuroshio ring E was pinched off from a meander of
the Kuroshio Extension jet on 10--12 June 2012 and disappeared in the middle of July. 
The eddy F has not been identified as a ring pinched off from the Kuroshio Extension.
The eddy G, located south-east of Hokkaido, is a typical warm-core ring of the Kuroshio.
It has been found to be closely related with the warm mesoscale eddy located to the south-west off
the Tohoku area (the Tohoku eddy T).

The frontal Kuroshio\,--\,Oyashio zone is populated with mesoscale eddies of different
sizes and lifetimes. They can be visualized on Lagrangian synoptic maps of the region
computed backward in time. We seed on a fixed day $1000 \times 1000$ tracers 
distributed homogeneously over the region shown on the maps and integrate them backward 
in time for 15 days starting from 28 June 2012. 
%The FTLE and drift maps in figs.3a, b and d have been computed backward 
%in time for 15 days starting from 28 June 2012. 
The mesoscale eddies are delineated in Fig.~\ref{fig3}a by
black ``ridges'' of the FTLE field which are known to approximate so-called
unstable manifolds of the hyperbolic trajectories \citep{H00} to be present in the area
during the integration period, 15 days in our case.
Backward-in-time drift map in Fig.~\ref{fig3}b on 28 June 2012 shows by shadows of grey color 
the absolute displacements of synthetic particles, $D$, in km for 15 days before the date indicated. 
The mesoscale eddies look like patches of different color than surrounding waters. 
The most prominent eddies with elliptic points in their centers
are also visible in the altimetric AVISO velocity field (Fig.~\ref{fig3}c). 

%The locations of the eddies have been also confirmed by computation of the absolute displacements
%of synthetic particles for the same period of time. The eddies on the corresponding
%drift maps, where shades of grey color code the displacements of particles from their initial positions to the final ones,
%look like patches of different color than surrounding waters (see Fig.~2s).

In the following numerical experiment tracers, distributed on 18 March 2011
over the box around the FNPP with the coordinates
(\E{140\text{--}144}; \N{36.5\text{--}38.5}), have been advected forward in time in the
altimetric velocity field.
In Fig.~\ref{fig4} we show the simulated density of tracers
during the cruise period in the North Pacific region, from 24 June to 6 July 2012.
The area with the Kuril mesoscale eddy A, centered at (\E{154.33}; \N{46.19}),
has not been visited during that period by potentially radioactive
tracers. The measured concentrations of \Cs{134} and \Cs{137} at station 50 in the center of
that eddy did not exceed the background level at 0, 200, 500 and 1000~m depth. As to the other
eddies of interest, the maximal radionuclide concentrations have been detected
in the centers of the eddies B, (centered at \E{154.4}; \N{41.9}), and G,
(centered at \E{147.3}; \N{41.3}), where
the simulated density of tracers is really higher in Fig.~\ref{fig4}.
The Kuroshio rings D, centered at  (\E{154}; \N{37.84}), and E, centered at
(\E{151.5}; \N{38.38}), look like
white patches on the map. It means that potentially radioactive tracers were
able to visit during the simulation period their periphery but not the core.

Besides of the eddies of interest, the increased density of tracers are clearly seen in Fig.~\ref{fig4}
at the places of the Tohoku eddy T with the center in the simulation period
at (\E{146.5}; \N{38}) and a cold-core cyclonic Kuroshio ring R, centered
at (\E{147.5}; \N{33}).
As to the Tohoku eddy, it was found by \citet{Kaeriyama13} to be
strongly contaminated by \Cs{134} and \Cs{137} just after the accident.
The increased simulated density
of tracers in the cold cyclonic ring R to the south of the Kuroshio Extension means that some
tracers from the initial patch could be transported southward to the parent jet and then be
trapped by that eddy. We have found that the ring R was
pinched off from a meander of the Kuroshio Extension jet and moved slowly
to the west along its southern
flank. In fact, it is a demonstration of transport of radionuclides across the strong Kuroshio
Extension jet documented and studied by \citet{Prants2014}.

Concluding this section, the reader is referred to Fig.~2s in supplementary material
where we show positions of those simulated tracers on 2 July 2012 
that have visited just after the accident (from 11 March to 10 April 2011)
two selected areas around the FNPP. The density of points in the centers of the eddies B, C and G,
where the highest cesium concentrations have been detected in the cruise, is really
comparatively large in this figure.

\subsection{Vertical structure of the mesoscale eddies and vertical
distribution of \Cs{134} and \Cs{137}}
\begin{figure}[!htb]
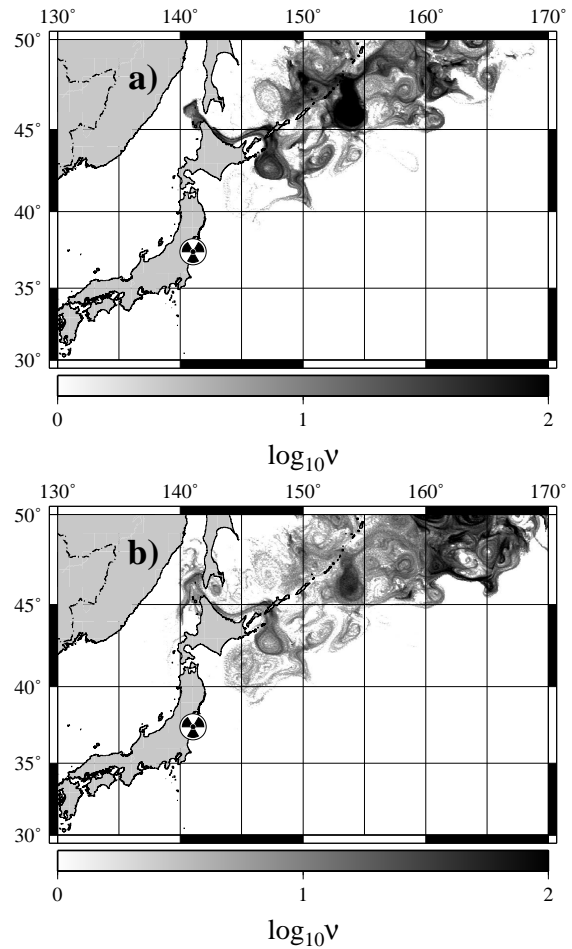

\centerline{\includegraphics[width=0.45\textwidth,clip]{fig8a.eps}}
\centerline{\includegraphics[width=0.45\textwidth,clip]{fig8b.eps}}
\caption{Backward-in-time tracking maps for tracers distributed (a) in
the center of the mesoscale eddy A
at station 50 and (b) at station 56, outside any eddy.
The maps show where the corresponding
tracers have been  just after the accident, from 11 March to 10 April 2011. 
The density of tracers, $\nu$, is in a logarithmic scale.}
\label{fig8}
\end{figure}
\begin{figure}[!htb]
\centerline{\includegraphics[width=0.45\textwidth,clip]{fig9a.eps}}
\centerline{\includegraphics[width=0.45\textwidth,clip]{fig9b.eps}}
\centerline{\includegraphics[width=0.45\textwidth,clip]{fig9c.eps}}
\caption{The same as in Fig.~\ref{fig8} but for the tracers distributed
in the center of (a) the mesoscale eddy B at station 61, (b) the mesoscale eddy C at station 65 and (c)
the mesoscale eddy D at station 69. The density of tracers, $\nu$, is in a logarithmic scale.}
\label{fig9}
\end{figure}
\begin{figure}[!htb]
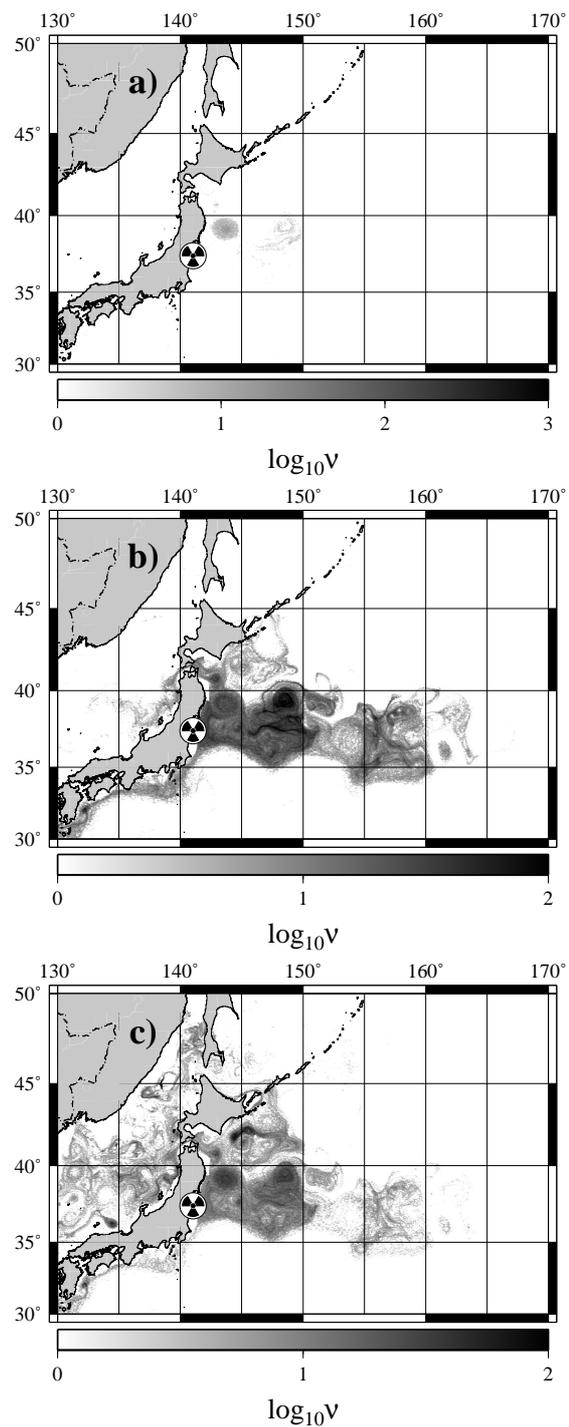

\centerline{\includegraphics[width=0.45\textwidth,clip]{fig10a.eps}}
\centerline{\includegraphics[width=0.45\textwidth,clip]{fig10b.eps}}
\centerline{\includegraphics[width=0.45\textwidth,clip]{fig10c.eps}}
\caption{The same as in Fig.~\ref{fig8} but for the tracers distributed
in the centers of (a) the mesoscale eddy E at station 74, (b) the mesoscale eddy F at station 79 and (c)
the mesoscale eddy G at station 84. The density of tracers, $\nu$, is in a logarithmic scale.}
\label{fig10}
\end{figure}

Results of CTD observations in the cruise demonstrate a very good correspondence 
with the modeled mesoscale eddy locations. Using the modeled maps of the eddy field
(Figs.~\ref{fig3}a--d) transmitted operationally on ship's board during the cruise, 
we managed to cross the
eddies very close to their centers. The only exception was the mesoscale eddy D that 
was crossed along its western
edge, not exactly through its center. Figure~\ref{fig5}a shows a vertical distribution
of potential density
anomaly along the cruise track (see Fig.~\ref{fig1}) from the central Kuril Islands in the north
(station 44) and down to the Kuroshio Extension in the south (station 70) and then to the northwest
toward Hokkaido (station 84).
As anticyclonic eddies contain water of lower density in their cores, their locations are clearly
seen by downward deflection of isopycnal lines. All of the 7 sampled eddies can be traced down
to 1000~m depth and five of them (A, B, D, F and G) can be seen down to the maximal depth of the CTD
observations (2000~m). Considering a magnitude of isopycnal lines deflection,  
the Kuril anticyclonic mesoscale 
eddy A seems to be the most intense dynamic feature among the sampled eddies. Isopycnes
of 26.8--27.6 are deepened in its center by 450--550~m as compared with surrounding water.
Another energetic eddies are the Kuroshio warm-core rings D and G.

Distribution of potential vorticity (Fig.~\ref{fig5}b) indicates an existence of low potential
vorticity
layers in the centers of eddies A, B, C and G in the depth range of 50--750~m. 
This corresponds to well
mixed vertically uniform cores of the eddies formed during eddy evolution
\citep{Kitano1975,Lobanov1991,Itoh2010}.
Vertical mixing, driven by winter convection, contributes
to the formation of eddy cores. Thus, the older eddies have larger and less stratified cores.
In opposite, relatively young eddies do not have a uniform layer in their centers. An absence of
any noticeable well mixed layer in the eddies D, E, and F confirms their relatively young age
detected by satellite altimetry and our modeling results. The relatively old Kuroshio
warm-core ring G demonstrated a more complicated structure having two low potential vorticity layers.
The upper one, corresponding to warm and higher salinity core of the eddy, is located between 55
and 205~m (Figs.~3s and 4s). While the secondary core of the eddy, formed by lower temperature 
and salinity waters subducted into the eddy center, is located at 410--750~m. 
Low potential vorticity cores of the eddies A, B, C and G
also have a high content of dissolved oxygen (Fig.~5s) which indicates 
recent ventilation of these waters. Considering these features of the eddy 
structure, we may expect accumulation of surface water with radionuclides in the centers of
anticyclonic eddies and vertical transport of this water downward in the eddy cores.

Distribution of \Cs{134} and \Cs{137} concentrations in surface waters
along the cruise track in the Pacific (Fig.~\ref{fig6})
shows increased cesium content in the areas of stations 56--65 and stations 81--84 corresponding
to the northern subarctic front areas along \E{155} and to the south-east off Hokkaido.
Maximal concentrations
are observed in the mesoscale eddy B located at the northern subarctic front with \Cs{134} and \Cs{137}
contents at station 61 up to 11.9 and 21.0~Bq~m$^{-3}$, respectively (Table 1).
High concentration of both the cesium isotopes were also
observed at the southern periphery of the mesoscale eddy B at station 63 (9.7 and 16.6~Bq~m$^{-3}$)
and in the area of the
mesoscale eddy G at station 84 (6.1 and 11.8~Bq~m$^{-3}$). This suggests a direct transport of 
water enriched by radionuclides from the Fukushima area by streamers and its trapping 
in those eddies that corresponds well with our modeling results.

The observed concentrations of cesium were low at the southern part of our transect 
(stations 70--79) within the surface layers of the eddies D, E, and
F, influenced by relatively clean water transported by the Kuroshio Extension. 
In the area to the north of the subarctic front and the central Kuril
Islands (the eddy A),
again we have not found a high content of radiocesium. Concentrations of
\Cs{134} and \Cs{137}  at stations
45--50 in the area of the eddy A were around 0.4--1.2 and 1.7--3.7~Bq~m$^{-3}$
which is higher than in the Japan
and Okhotsk Seas. However, this is rather a result of atmospheric deposition then direct advection
by water flow.

Vertical distribution of radiocesium, sampled at some stations, is shown in Fig.~\ref{fig7}.
Inside most of
the eddies a higher concentration is observed not at the surface but withim subsurface layers deepened
down to 200--500~m. The maximal content of cesium isotopes in the eddies B, C and G
(stations 61, 65 and 84)
is observed in the low vorticity cores of these eddies located around 200, 200 and 100~m, correspondingly
(Fig.~\ref{fig5}b). This proves our preliminary hypothesis that water with Fukushima-derived 
radionuclides
was subducted and trapped in the cores of anticyclonic eddies. Lower content of cesium at surface
layer of the eddies may be explained by a faster ventilation of surface layer by mesoscale streamers
and advection of relatively clean water originated from the Kuroshio Extension.

Our observations show that maximal concentrations of cesium in the subarctic frontal zone in
June\,--\,July 2012 was located not at the surface but within subsurface and intermediate 
water layers in the potential density range of 26.5--26.7~$\sigma_{\theta}$.
Deepening of these isopycnal surfaces in anticyclonic
eddies resulted in deeper locations of radionuclide maxima. Thus, high concentrations of cesium
were observed down to 300--400~m in the eddies F, E and G (stations 75, 79 and 84) and down to 500~m
in the mesoscale eddy D (station 69). Even deeper penetration of \Cs{134} and \Cs{137},
down to 1000~m, was observed
in the mesoscale eddy B (station 61) of $0.8 \pm 0.2$ and $1.9 \pm 0.2$~Bq~m$^{-3}$.

The first observations of vertical distribution
of Fukushima-derived cesium, taken in summer of 2011 \citep[e.g.,][]{Buesseler12}, found increased
content of cesium isotope at the upper surface layer down to 50--100~m. 
Then, starting from spring 2012 and later, a higher concentration of
cesium was observed subducted down to subsurface and intermediate layers
\citep{Kaeriyama2014,Kumamoto2014}. This coincides well with
our results which show that subduction in the frontal zone forms a
cesium enriched intermediate water which then spreads southward at the
depth of 200--370~m between potential density anomalies surfaces of 25.20--25.33 
as North Pacific Subtropical Mode Water \citep{Kumamoto2014}.
Modeling results by \citep{Rossi2013} also demonstrate deep penetration
of radiocesium rich water into the North Pacific Mode Water in the eastern
areas of the North Pacific. We demonstrate that on the background of
this large-scale subduction and advection anticyclonic mesoscale eddies
could be considered as an effective mechanism of downward transport and
advection of cesium-rich water. Our observations prove that high radiocesium 
concentrations are observed exactly inside the uniform core of the
eddies (termostad) located in the density range 26.5--26.7 which corresponds to the 
North Pacific Intermediate Water forming in the subarctic frontal zone
area \citep{Talley1995,Shimizu}. Thus, we expect that
the eddies can trap this water with the maximal content of radiocesium and
transport it horizontally and vertically. Considering the northward
translation of the eddies, we may suggest that Fukushima-derived cesium
should be also transported by the eddies to the north at the intermediate depth. 

\subsection{Backward-in-time tracking maps for samples collected at stations
in the centers of anticyclonic eddies of the subarctic front}

In this section we present the results of numerical simulation with tracers
distributed in the centers of the eddies A, B, C, D, E, F and G where seawater
samples were collected at some cruise stations. 
All the tracking experiments here have been performed backward in time for each mesoscale 
eddy to be studied. Starting from the day of sampling in the center of the corresponding eddy, 
synthetic tracers have been advected backward in time till the day of the tsunami, 11 March 2011. 
Fixing the places, where they have been  
for the month after the accident, from 11 March to 10 April 2011, we plot the corresponding map. 

Station 50 (\E{154.33}; \N{46.19}) was located
near the elliptic point of the mesoscale eddy A, which is a Kuril anticyclone with the size
$\simeq 3^\circ \times 1.5^\circ$ located
approximately at the same place from the day of the accident (and even earlier)
to the end of the cruise (and even later). The observed concentrations
at different depths (see Table) ranged for \Cs{137} from $0.6 \pm 0.1$
to $1.7 \pm 0.3$~Bq~m$^{-3}$ and did not exceed the background level.
It is seen in Fig.~\ref{fig8}a that the tracers of station
50 have not visited the latitudes to the south off \N{40} where one would expect 
a significant contamination due to the accident. We may
conclude that the probability to detect an increased cesium concentration in the eddy A
is small.  For comparison, we computed in Fig.~\ref{fig8}b a tracking map for tracers 
of station 56
(\E{155}; \N{44.05}) located outside any eddy. The observed concentration of
\Cs{137} in surface water samples at that station, $5.9 \pm 0.5$~Bq~m$^{-3}$,
exceeds the background level more than in three times. The traces of simulated particles 
in Fig.~\ref{fig8}b
are found to be closer to the FNPP location than the tracers in Fig.~\ref{fig8}a 
(please, pay attention that the concentration in those figures is in a logarithmic scale).
%In simulation we distribute a large number of tracers around location
%of the station of interest and advect them in the altimetric
%velocity field backward in time to
%the day of the accident. Then we fix on the corresponding tracking map the places
%where those tracers have been  for one month after the accident, from 11 March to 10 April 2011.

Station 61 (\E{154.4}; \N{41.9}) was located
near the elliptic point of the mesoscale eddy B with the size $\simeq 1.5^\circ \times 1.5^\circ$.
The highest cesium concentrations, $21.1 \pm 1.1$ at surface and
$21.6 \pm 0.9$~Bq~m$^{-3}$ at 203~m depth,
have been observed in seawater samples at that station. Our observations are
agreed with measurements by \citet{Kaeriyama13} to be carried out approximately at the same place
and in the same period.
They detected the concentration of \Cs{137} in surface seawater samples, $18 \pm 0.7$~Bq~m$^{-3}$,
at their station B38 located nearby our station 61 and
$17 \pm 0.7$~Bq~m$^{-3}$ and
$13 \pm 0.7$~Bq~m$^{-3}$ at stations B37 and B39 located inside the eddy B.
The tracking map in Fig.~\ref{fig9}a shows that tracers of the eddy B have visited for the month
after the accident the area with presumably high level of contamination. In particular,
they have visited
frequently the location of the Tohoku eddy T in March and April 2011 with the highest levels
of cesium concentration (besides the FNPP discharge channels) to be detected just after
the accident \citep{Kaeriyama13}.
We traced out the history of the mesoscale eddy B and found that it was born on the southern
flank of a zonal eastward jet transporting waters from the eastern coast of
Japan. That jet is seen on the computed zonal drift Lagrangian map in Fig.~\ref{fig3}d as
a wave-like grey (red online) band extending approximately along \N{42\text{--}43}.
Red (blue) colors mean that the corresponding tracers passed for the integration time (15 days)
large distances in the west\,--\,east and the east\,--\,west directions, respectively.

Station 65 (\E{154}; \N{39.75}) was located
near the elliptic point of the mesoscale eddy C with the size $\simeq 1^\circ \times 1^\circ$,
which was born in January 2012 as a companion of the warm-core Kuroshio Extension ring D.
The high concentration of \Cs{137}, $22.7 \pm 1.5$~Bq~m$^{-3}$, has been
detected at 200~m depth,
whereas it was relatively low at the surface and at 500 and 1000~m depths
(see Table).
The concentrations of \Cs{137} in surface seawater samples at Japanese stations A34
and  A33, located nearby our station 65, and at station A35 located
in the core of the eddy C were
found by \citep{Kaeriyama13} to be relatively high, $9.2 \pm 0.5$, $11 \pm 0.6$ and
$13 \pm 0.7$~Bq~m$^{-3}$. The probability that waters of the eddy C could
contain a large amount of
Fukushima-derived radionuclides is comparatively high because the tracers in Fig.~\ref{fig9}b
have frequently visited the contaminated area and, in particular, the Tohoku eddy T.

The Kuroshio ring D with the size $\simeq 2.5^\circ \times 3^\circ$ was pinched off
from a meander of the Kuroshio Extension jet in the end of May 2012. Until the middle of August it
was a free ring sometimes to be connected with the parent jet by an arch. The probability to detect
higher concentrations of cesium in its surface water is estimated to be low (see Fig.~\ref{fig9}c) because
it contains mainly Kuroshio waters. We have detected in surface water samples the concentration
of \Cs{137} to be $6.3 \pm 0.4$~Bq~m$^{-3}$. It is greater than the background level that
may be explained by water exchange with its companion, the mesoscale eddy C with a high level 
of radioactivity. The concentrations of \Cs{137}
in surface seawater samples at Japanese stations B30, located closely to our station 69, and
at station B29, located nearby our station 70, were
found by \citep{Kaeriyama13} to be slightly greater the background level, $3.6 \pm 0.5$ and
$3.4 \pm 0.4$~Bq~m$^{-3}$, respectively.

The Kuroshio ring E with the size $\simeq 1.5^\circ \times 1^\circ$ was pinched off
from a meander of the jet on 10--12 June 2012 and disappeared in the middle of July.
Station 74 (\E{151.5}; \N{38.38}) was located
near the elliptic point of that eddy where the increased concentration
of \Cs{137}, $12.3 \pm 0.8$~Bq~m$^{-3}$, has been detected at 307~m depth.
A comparatively small number of tracers over the whole broad area in Fig.~\ref{fig10}a
is explained by the history of core waters in the mesoscale eddy E which have been transported
mainly by the Kuroshio from the south and then directed to the east by the
Kuroshio Extension. The genesis of the eddy E
shows a presence of the Tohoku eddy waters in its core
(see the patch in Fig.~\ref{fig10}a centered at \E{{\approx}143}; \N{39}).

The mesoscale eddy F with the size $\simeq 1^\circ \times 1^\circ$ has not been identified as
a ring pinched off from the Kuroshio Extension. Station 79 (\E{149.5}; \N{39.5})
was located near the elliptic point of that eddy where the increased concentrations of \Cs{137},
ranged from $7.3 \pm 0.7$ to $9 \pm 0.9$~Bq~m$^{-3}$, have been detected
from the surface to 280~m depth. The tracking map in Fig.~\ref{fig10}b demonstrates that water
from its core really have visited potentially contaminated area around the FNPP location
during the first month after the accident.

Station 84 (\E{147.3}; \N{41.3})
was located near the elliptic point of the mesoscale eddy G with the size
$\simeq 2^\circ \times 1.5^\circ$
situated at the traverse of the Tsugaru Strait. The tracking map for
that station in Fig.~\ref{fig10}c reveals
its close connection with the Tohoku eddy T, and, therefore, the probability to detect increased
cesium concentrations was expected to be comparatively large. In reality we detected the
concentration of \Cs{137} at 100~m depth to be as largr as $18 \pm 1.3$~Bq~m$^{-3}$.

\section{Conclusion}

Results of observations of \Cs{134}  and \Cs{137}, released from the Fukushima Nuclear Power Plant 
in seawater samples and collected at surface and different depths in the western North Pacific 
in June and July 2012 in the cruise of R/V ``Professor Gagarinskiy'', have demonstrated a background 
or slightly increased level of radiocesium in the Japan and Okhotsk seas and increased concentrations 
in the area of the subarctic front east of Japan. The highest concentrations of \Cs{134}  and \Cs{137}
(13.5 ${\pm}$ 0.9 and 22.7 ${\pm}$ 1.5 Bq~m$^{-3}$) have been found to 
exceed ten times the background levels before the accident.  
Maximal content of radiocesium was observed inside the mesoscale anticyclonic eddies. 
Among the sampled eddies, the anticyclonic ring of the northern subarctic front B and the  
Kuroshio warm-core ring G (located south-east of Hokkaido) presented the highest concentration 
of cesium. 

The maximal  
concentration of radionuclides was observed not at the surface but within subsurface and intermediate 
water layers (100 -- 500~m) in the potential density range of 26.5 -- 26.7 with extremely high 
values inside the low potential vorticity cores of the eddies B, C and G. 
This suggests that convergence 
and subduction of surface water inside eddies are main mechanisms of downward transport of 
radionuclides. In particular, in the eddies B and D a slightly increased content of radiocesium was 
observed even at depth of 1000~m. Concentrations of radiocesium in the eddies, located closer 
to the Kuroshio 
Extension (D, E and F), have been found to be very low at surface layer. It may be explained 
by a faster ventilation of surface layer by mesoscale streamers and advection of relatively 
clean water originated 
from the Kuroshio Extension. A higher content of radiocesium observed at 300--500~m depth was 
a result of sudbuction in the frontal zone and the following advection of intermediate water by 
the eddies. 

The direct observations were compared with simulation of advection of these 
radioisotopes by the AVISO altimetric velocity field. Synthetic tracers, released at the locations 
of a number of stations inside the eddies, have been advected backward in time till the day of the 
accident. Fixing their traces for the month after the accident, we computed tracking maps 
for each of those stations which were used to reconstruct the history and origin of the tracers 
imitating measured seawater samples. Those maps allowed to explain why measured activities of 
\Cs{134}  and \Cs{137} differed strongly in different samples. It has been shown that tracers with increased 
radioactivity really have visited the areas with presumably high level of contamination just after 
the accident. In particular, it has been found a close connection of the samples with increased 
radioactivity in the eddies B, C, E, F and G with the Tohoku eddy T  known to be strongly contaminated 
just after the accident \citep{Honda12,Buesseler12,Kaeriyama13}. 

Thus, the eddies play an important role in the transport of water with radionuclides released from 
the FNPP. Considering subduction and accumulation of high-cesium-content water in the anticyclonic 
eddies, we may suggest that Fukushima-derived cesium should be also transported by the eddies 
northward at the intermediate depth.

Supplementary material related to this paper is available online.

\section*{Acknowledgments}

The authors would like to thank the research team and crew of R/V ``Professor Gagarinskiy''
for their hard work at sea. The cruise was supported by the Far Eastern Branch of Russian Academy
of Sciences and the Russian Foundation for Basic Research (the expedition grant~11--05--02110).
The modeling work was supported  by the Russian Foundation for Basic Research
(project nos.~12--05--00452, 13--05--00099 and 13--01--12404).
The altimeter products were distributed by AVISO with support from CNES.

%\section*{Appendix}
%\afterpage{\include{tab}}
\onecolumn
\tabulinesep = _1mm^1mm
{%\footnotesize
\small\normalfont
%\begin{longtabu} to \textwidth {|X[0.9,c]|X[1.4,c]|X[1,c]|X[0.9,c]|X[0.95,r]|X[0.95,r]|X[0.6,r]|X[0.9,c]|X[0.9,c]|}
\begin{longtabu} to \textwidth {X[0.9,c]X[1.4,c]X[1,c]X[0.9,c]X[0.95,r]X[0.95,r]X[0.6,r]X[0.9,c]X[0.9,c]}
\hline\\
\rowfont[c]{}
Station&Time&Longitude&Latitude&Depth,~m&Temp,~$^\circ$C&Sal, psu&\Cs{134}, Bq/m$^3$&\Cs{137}, Bq/m$^3$\\
\hline\endhead
\hline\endfoot
\taburowcolors{white .. gray!20}
1      & 12.06.12~13:50 & \E{131.877} & \N{43.067} & 0  &$12.401$&32.117& $<0.09$    &$1.9\pm0.2$ \\%\showrowcolors
7      & 14.06.12 01:56 & \E{131.995} & \N{42.514} & 0  & $9.700$&33.860& $<0.08$    &$1.7\pm0.2$ \\
11     & 14.06.12 11:55 & \E{132.004} & \N{42.220} & 0  &$10.390$&33.850& $<0.08$    &$1.9\pm0.3$ \\
16     & 15.06.12 14:18 & \E{132.517} & \N{40.668} & 0  &$15.198$&33.972& $<0.05$    &$1.4\pm0.2$ \\
       &                &             &            & 201& $1.454$&33.978& $<0.07$    &$1.7\pm0.3$ \\
       &                &             &            & 751& $0.533$&34.069& $<0.07$    &$1.7\pm0.3$ \\
       &                &             &            &1500& $0.195$&34.066& $<0.08$    &$0.7\pm0.1$ \\
       &                &             &            &2500& $0.110$&34.065& $<0.06$    &$0.4\pm0.1$ \\
       &                &             &            &3361& $0.090$&34.065& $<0.09$    &$1.0\pm0.3$ \\
21     & 16.06.12 21:21 & \E{133.167} & \N{40.003} & 0  &$17.607$&34.000& $<0.07$    &$2.2\pm0.3$ \\
24     & 17.06.12 19:09 & \E{133.997} & \N{41.504} & 0  &$14.609$&33.950& $<0.08$    &$1.6\pm0.2$ \\
25     & 18.06.12 05:01 & \E{135.121} & \N{41.520} & 0  &$14.370$&33.880& $<0.05$    &$2.0\pm0.2$ \\
       &                &             &            &2900& $0.087$&34.064& $<0.06$    &$0.3\pm0.1$ \\
26     & 18.06.12 12:00 & \E{135.722} & \N{41.847} & 0  &$14.657$&33.852& $<0.08$    &$1.6\pm0.2$ \\
27     & 19.06.12 07:08 & \E{137.183} & \N{42.818} & 0  &$15.212$&33.908& $<0.07$    &$2.2\pm0.2$ \\
       &                &             &            &3621& $0.088$&34.065& $<0.06$    &$1.1\pm0.1$ \\
28     & 19.06.12 15:15 & \E{137.842} & \N{43.501} & 0  &$15.690$&33.967& $<0.09$    &$2.3\pm0.3$ \\
29     & 20.06.12 01:21 & \E{138.696} & \N{44.311} & 0  &$15.306$&34.030& $<0.08$    &$2.0\pm0.2$ \\
L2     & 21.06.12 01:30 & \E{141.408} & \N{45.721} & 0  &$11.980$&33.806&$0.3\pm0.2$ &$2.4\pm0.3$ \\
L3     & 21.06.12 08:00 & \E{142.110} & \N{45.686} & 0  & $9.045$&33.540& $<0.08$    &$1.8\pm0.2$ \\
L4     & 21.06.12 14:00 & \E{142.935} & \N{45.737} & 0  & $8.918$&32.203& $<0.06$    &$1.5\pm0.2$ \\
L5     & 21.06.12 20:30 & \E{144.024} & \N{45.805} & 0  & $8.355$&32.242& $<0.06$    &$1.5\pm0.2$ \\
30     & 22.06.12 08:16 & \E{146.076} & \N{45.934} & 0  & $8.206$&32.281& $<0.03$    &$1.5\pm0.1$ \\
       &                &             &            & 150&$-0.340$&33.180& $<0.07$    &$1.4\pm0.1$ \\
       &                &             &            & 500& $1.560$&33.640& $<0.03$    &$1.1\pm0.1$ \\
       &                &             &            &1005& $2.220$&34.240& $<0.07$    &$0.9\pm0.2$ \\
       &                &             &            &2200& $1.785$&34.506& $<0.03$    &$0.7\pm0.1$ \\
       &                &             &            &3236& $1.626$&34.604& $<0.07$    &$0.4\pm0.1$ \\
31     & 23.06.12 07:43 & \E{148.393} & \N{46.456} & 0  & $8.591$&32.272& $<0.05$    &$1.6\pm0.1$ \\
35     & 24.06.12 05:36 & \E{150.932} & \N{47.043} & 0  & $5.810$&32.640& $<0.06$    &$1.5\pm0.1$ \\
43     & 24.06.12 23:11 & \E{151.609} & \N{46.204} & 0  & $3.843$&32.976&$0.4\pm0.1$ &$1.9\pm0.5$ \\
44     & 25.06.12 10:57 & \E{152.732} & \N{47.050} & 0  & $6.010$&32.910&$0.9\pm0.2$ &$2.6\pm0.2$ \\
45     & 25.06.12 15:16 & \E{153.095} & \N{46.803} & 0  &        &      &$1.2\pm0.2$ &$3.7\pm0.4$ \\
       &                &             &            & 200&        &      &$0.4\pm0.1$ &$1.9\pm0.2$ \\
       &                &             &            & 500&        &      & $<0.07$    &$0.9\pm0.1$ \\
       &                &             &            &1000&        &      & $<0.08$    &$0.7\pm0.1$ \\
49     & 26.06.12 12:10 & \E{154.577} & \N{46.083} & 0  & $6.700$&33.030&$0.4\pm0.1$ &$1.8\pm0.2$ \\
50     & 26.06.12 14:54 & \E{154.334} & \N{46.197} & 200& $1.320$&33.360&$0.5\pm0.2$ &$1.7\pm0.3$ \\
       &                &             &            & 500& $1.680$&33.460&$0.2\pm0.1$ &$1.5\pm0.1$ \\
       &                &             &            &1000& $3.240$&34.150& $<0.03$    &$0.6\pm0.1$ \\
54     & 27.06.12 08:03 & \E{155.279} & \N{45.717} & 0  & $6.501$&32.932&$1.2\pm0.1$ &$3.0\pm0.3$ \\
56     & 27.06.12 23:05 & \E{155.000} & \N{44.050} & 0  & $7.256$&33.011&$3.0\pm0.3$ &$5.9\pm0.5$ \\
       &                &             &            & 200& $2.932$&33.573&$0.5\pm0.2$ &$2.0\pm0.2$ \\
       &                &             &            & 500& $3.334$&34.122& $<0.09$    &$0.9\pm0.1$ \\
       &                &             &            &1000& $2.554$&34.404& $<0.07$    &$0.5\pm0.1$ \\
       &                &             &            &2002& $1.705$&34.601& $<0.06$    &$0.4\pm0.1$ \\
       &                &             &            &3002& $1.297$&34.661& $<0.09$    &$0.5\pm0.3$ \\
58     & 28.06.12 17:27 & \E{155.005} & \N{42.999} & 0  &$12.800$&34.020&$2.4\pm0.2$ &$5.2\pm0.4$ \\
61     & 29.06.12 06:30 & \E{154.438} & \N{41.936} & 0  &$11.580$&33.985&$11.9\pm0.6$&$21.0\pm1.1$\\
       &                &             &            & 203&$ 6.817$&33.904&$12.5\pm0.5$&$21.6\pm0.9$\\
       &                &             &            & 500&$ 4.135$&33.907&$0.2\pm0.1$ &$1.6\pm0.3$ \\
       &                &             &            &1000&$ 3.064$&34.347&$0.8\pm0.2$ &$1.9\pm0.2$ \\
63     & 29.06.12 18:15 & \E{154.262} & \N{41.068} & 0  &$12.100$&33.940&$9.7\pm0.5$ &$16.6\pm0.9$\\
65     & 30.06.12 06:42 & \E{154.003} & \N{39.750} & 0  &$17.139$&34.444&$2.1\pm0.2$ &$4.8\pm0.3$ \\
       &                &             &            & 200& $9.218$&34.155&$13.5\pm0.9$&$22.7\pm1.5$\\
       &                &             &            & 500& $5.017$&33.926& $<0.07$    &$0.9\pm0.2$ \\
       &                &             &            &1000& $3.173$&34.354& $<0.11  $  &$0.7\pm0.2$ \\
69     & 01.07.12 08:44 & \E{154.001} & \N{37.842} & 0  &$18.650$&34.690&$2.8\pm0.2$ &$6.3\pm0.4$ \\
       &                &             &            & 205&$15.990$&34.630&$4.1\pm0.3$ &$8.0\pm0.5$ \\
       &                &             &            & 500& $7.770$&34.030&$3.5\pm0.3$ &$7.2\pm0.6$ \\
       &                &             &            &1000&        &      & $<0.07$    &$1.1\pm0.2$ \\
70     & 01.07.12 20:25 & \E{153.981} & \N{37.366} & 0  &$22.368$&34.390&$0.2\pm0.1$ &$2.3\pm0.2$ \\
71     & 02.07.12 06:15 & \E{152.813} & \N{37.863} & 0  &$18.304$&34.313&$1.6\pm0.2$ &$4.5\pm0.3$ \\
74     & 02.07.12 19:24 & \E{151.536} & \N{38.385} & 0  &$19.120$&34.518&$0.6\pm0.1$ &$2.5\pm0.2$ \\
       &                &             &            & 100&$14.860$&34.530&$1.1\pm0.2$ &$4.0\pm0.4$ \\
       &                &             &            & 307& $5.893$&33.885&$5.6\pm0.6$ &$12.3\pm0.8$\\
       &                &             &            & 500& $4.720$&34.021& $<0.07$    &$1.4\pm0.1$ \\
76     & 03.07.12 09:56 & \E{150.701} & \N{38.836} & 0  &$18.650$&34.235& $<0.06$    &$0.4\pm0.1$ \\
77     & 03.07.12 16:37 & \E{150.002} & \N{39.151} & 0  &$18.290$&34.370&$1.5\pm0.2$ &$4.1\pm0.3$ \\
79     & 04.07.12 02:41 & \E{149.003} & \N{39.495} & 0  &$18.174$&34.319&$1.7\pm0.2$ &$4.3\pm0.3$ \\
       &                &             &            & 100&$10.643$&34.236&$2.9\pm0.4$ &$7.3\pm0.7$ \\
       &                &             &            & 280& $3.906$&33.489&$3.5\pm0.6$ &$9.0\pm0.9$ \\
       &                &             &            & 500& $4.229$&33.964& $<0.05$    &$1.3\pm0.2$ \\
81     & 04.07.12 14:54 & \E{148.387} & \N{40.234} & 0  &$15.950$&33.530&$3.1\pm0.3$ &$6.7\pm0.5$ \\
82     & 04.07.12 20:20 & \E{147.996} & \N{40.631} & 0  &$13.625$&33.242&$1.0\pm0.1$ &$2.5\pm0.2$ \\
84     & 05.07.12 05:29 & \E{147.329} & \N{41.299} & 0  &$14.750$&33.846&$6.1\pm0.4$ &$11.0\pm0.6$\\
       &                &             &            & 100& $8.352$&33.979&$10.4\pm0.7$&$18.0\pm1.3$\\
       &                &             &            & 350& $4.672$&33.601&$6.9\pm0.4$ &$11.3\pm0.6$\\
       &                &             &            & 500& $1.895$&33.432&$0.4\pm0.1$ &$1.7\pm0.2$ \\
\taburowcolors{white .. white}
\hline\\
\caption{Concentrations of \Cs{134} and \Cs{137} in seawater collected 
in the western North Pacific including the Japan and Okhotsk Seas.}
\end{longtabu}
}
\twocolumn

\bibliography{elsarticle-template-2-harv}
\bibliographystyle{model2-names}
\end{document}